\documentclass[12pt,a4paper]{article}
\begin{document}
\baselineskip=1.5\baselineskip
\title{Linear aggregation beyond isodesmic symmetry} 
\author{J. R. Henderson\\ 
\\
School of Physics and Astronomy\\
and Centre for Self-Organising Molecular Systems\\ 
University of Leeds, Leeds LS2 9JT, UK}
\date{\today}
\maketitle
\newpage
\section*{Abstract}
Exactly solvable models of linear aggregation have been known since Ising's seminal one-dimensional model. This model is defined by a unique nearest-neighbour bond strength that is independent of the length of the cluster; known as isodesmic symmetry. Linear aggregation in real systems has often been associated with broken isodesmic symmetry. Here we show that important examples can be mapped to a class of one-dimensional models that are also exactly solvable.

\newpage
\section*{}
A model has isodesmic symmetry if every member of the infinite set of cluster equilibria $[1]+[n-1]\Leftrightarrow [n]$ define the same equilibrium `constant', as is appropriate to linear aggregation (self-assembly) where the energy required to break a chain aggregate is independent of the position of the break and the length of the chain. The one dimensional (1D) Ising \cite{ising} and equivalent Lattice-Gas (LG) models are isodesmic everywhere in phase-space. We shall consider the generalisation where the first member of the set only, $[1]+[1]\Leftrightarrow [2]$, has a different equilibrium constant to all the rest. When treated phenomenologically with the ``law of mass action'' as an implicit solvent model this is a well-known problem in the physical chemistry of aggregation \cite{PCagg} and is also the basis of current understanding of the self-assembly of peptide tapes involved in protein miss-folding diseases such as Alzheimer's dementia \cite{peptides}.\\

Let us specify a 1D LG hamiltonian with two energies; namely, an energy $\epsilon_m$ that an isolated monomer $[1]$ must take up in the process of bonding to a cluster, and an attractive nearest-neighbour energy $\epsilon+\epsilon_m$ for each bond within a cluster of length $n>1$. Then, when a monomer joins a cluster $[n>1]$ the change in energy involved is $\epsilon$, but when two monomers combine to form a dimer the gain in attractive energy is only $\epsilon-\epsilon_m$. Let us use the exponential factor $\tau\equiv e^{\beta\epsilon}$ to represent a temperature field ($\beta$ denotes inverse temperature in units of Boltzmann's constant) and it is convenient to define $\gamma\equiv \epsilon_m/\epsilon$. A direct route to the exact equation of state and cluster distribution is to apply Widom's potential distribution theorem (PDT), \cite{widom}:
\begin{equation} 
f\equiv \frac{\rho}{1-\rho} = z\langle e^{-\beta\psi(x)}\rangle_c\,\,,
\label{pdt}
\end{equation}
where $z=e^{\beta\mu}$ denotes the monomer activity ($\mu$ is chemical potential), $\rho$ is the density in LG units (probability of occupancy of a single cell), the subscript $c$ indicates a conditional average (there is an empty cell at position $x$ enforced by the $1/(1-\rho$) factor on the left side of the equation) and $\psi$ denotes the change in energy that would arise if a monomer were to be inserted into this empty cell. The evaluation of this average (or ratio of partition functions) can be broken down into insertions into a finite set of specific clusters; let us use $\rho_w$, $f_w$ for occupancy next to an empty cell, $\rho_1$, $f_1$ for occupancy next to a monomer isolated on the other side and $\rho_a$, $f_a$ for occupancy next to the end of an fixed aggregate (needed below to evaluate the cluster distribution). Figure~1 enumerates all the possibilities required for our model, together with the corresponding values of $\psi$. One can then use conditional probability to evaluate (\ref{pdt}) by inspection (this can be done in any order but here I shall start at the insertion cell and move right and left):
\begin{eqnarray}
f&=&z\left\{(1-\rho_w)^2+2(1-\rho_w)\rho_w(1-\rho_1)\tau^{1-\gamma}
+2(1-\rho_w)\rho_w\rho_1\tau \right.\nonumber\\
&+&\rho_w^2(1-\rho_1)^2\tau^{2-\gamma}+2\rho_w^2(1-\rho_1)\rho_1\tau^2
+\rho_w^2\rho_1^2\tau^{2+\gamma}\left.\right\}\\
&=& z\left\{(1-\rho_w)^2+2(1-\rho_w)\left[\rho_w(1-\rho_1)\tau^{1-\gamma}+\rho_w\rho_1\tau\right]\right. \nonumber\\
&+& \tau^{\gamma}\left[\rho_w(1-\rho_1)\tau^{1-\gamma}+\rho_w\rho_1\tau\right]^2 
\left.\right\} \,\,.
\label{eos1}
\end{eqnarray}
To evaluate $f_w$, $f_1$ (and $f_a$), one simply applies the PDT to insertions in the presence of the fixed clusters at the left of the insertion cells depicted in Fig.~2; each case is associated with three non-degenerate states. From Fig.~2a we have 
\begin{equation}
f_w=z\left\{(1-\rho_w)+\left[\rho_w(1-\rho_1)\tau^{1-\gamma}+\rho_w\rho_1\tau\right]\right\}\\
\,\,,
\label{fw}
\end{equation} 
while from Fig.~2b and Fig.~2c, respectively,
\begin{eqnarray}
f_1&=&f_a\tau^{-\gamma}\,\,,\label{fa}\\
f_a&=&z\left\{(1-\rho_w)\tau+\rho_w(1-\rho_1)\tau^{2}+\rho_w\rho_1\tau^{2+\gamma}\right\}\,\,.
\label{f1fa}
\end{eqnarray}
Inserting (\ref{fw}) into (\ref{eos1}) leads to a simplification:
\begin{equation}
f=\frac{f_w^2}{z}+(\tau^{\gamma}-1)\left[\frac{f_w}{\sqrt{z}}-\frac{\sqrt{z}}{1+f_w}\right]^2\,\,.
\label{eos}
\end{equation}
To evaluate $f_w(\tau,z,\gamma)$ and hence also obtain $f$, $f_1$ and $f_a$, one can solve (\ref{fw}-\ref{f1fa}) as a pair of simultaneous equations; e.g. 
\begin{eqnarray}
f_w(1+f_w)(1+f_1) &=& z(1+f_1+f_w\tau^{1-\gamma}+f_1f_w\tau)\\
f_w(1+f_w)\tau-f_1(1+f_w) &=& z(\tau-\tau^{1-\gamma})\,\,.
\end{eqnarray}
The solution yields 
\begin{eqnarray}
f_1(1+f_1) &=& f_w(\tau^{1-\gamma}+f_1\tau)\,\,,\label{f1fw}\\
f_w(1+f_w) &=& z(1+f_1) \,\,,
\label{fwf1}
\end{eqnarray}
together with a cubic equation of state for $f_w$:
\begin{equation}
z^2(\tau-\tau^{1-\gamma})=z(1+f_w)(1+f_w\tau)-f_w(1+f_w)^2
\,\,.
\label{cubic}
\end{equation}
Although the general mathematical solution for one-dimensional models is invariably a cubic, for short-range models in equilibrium there must never be more than one physical solution in order to preserve the absence of phase transitions in one-dimension \cite{LL}. For our non-isodesmic model we see that this arises because (\ref{cubic}) is also a quadratic for the activity, which only ever has one physical solution:
\begin{equation}
2z(\tau-\tau^{1-\gamma})=(1+f_w)\left[1+f_w\tau-\sqrt{(1+f_w\tau)^2-4f_w(\tau-\tau^{1-\gamma})}\right]
\,\,.\label{soln}
\end{equation} 
In this version of the equation of state the variables $\tau$ and $f_w$ are a field and a density, respectively. One can transform the equation of state into $\mu(\tau,\rho)$ by using $f_w$ as an implicit variable varying between zero and infinity in both equations (\ref{soln}) and (\ref{eos}). The occupation variables $f_1$ and $f_a$ similarly follow from (\ref{fwf1}) and (\ref{fa}), respectively. Note that there is no region of phase space associated with imaginary solutions contained within the above analysis.\\

The entire cluster distribution is also defined by the above occupation probabilities: 
\begin{eqnarray}
[1] &=& (1-\rho)\rho_w(1-\rho_1) \nonumber\\
&=& (1-\rho)(1-\rho_w)\rho_a\tau^{-1} \,\,,\label{one}\\
\left[n>1\right] &=& (1-\rho)\rho_w\rho_1\rho_a^{n-2}(1-\rho_a)\nonumber\\
&=& [1]\tau^{-\gamma}\rho_a^{n-1}\,\,.
\label{cdis}
\end{eqnarray}
Here, I have chosen to build the clusters in strict order of left to right and note that an isolated monomer is actually a cluster of two empty sites enclosing an occupied site. Note also that to obtain the second line of (\ref{one}) one requires eqn.~(\ref{f1fw}). It is straightforward to directly check the self-consistency of the PDT by showing that (\ref{eos}) is also the solution of the sum rule
\begin{equation}
\rho = [1] + \sum_{n>1}\,\,n[n] \,\,.
\label{srrho}
\end{equation}
The clusters $[n>1]$ satisfy the isodesmic exponential distribution $[n>1]=[1]\tau^{-\gamma}\rho_a^{n-1}$, renormalized by a factor $\tau^{-\gamma}$, and thus have an average cluster length
\begin{equation}
\bar{N}_t\equiv\frac{\sum_{n>1}n[n]}{\sum_{n>1}[n]} = 2+f_a = 2+f_1\tau^{\gamma}\,\,. 
\label{nbar}
\end{equation}
The only effect of the broken isodesmic symmetry is to cause the aggregation distribution to step down a factor of $\tau^{-\gamma}$ between isolated monomers ($n=1$) and dimers ($n=2$). In the case of peptide tape formation, the broken isodesmic symmetry ($\gamma>0$) results from the energy barrier that must be overcome by a helical monomer when unwinding in the process of joining a beta-sheet tape. Experimental procedures exist that can directly measure the percentage of peptide in a beta-sheet confirmation, \cite{peptides}. This defines the order parameter
\begin{equation}
\phi_t \equiv \frac{1}{\rho}\sum_{n>1}n[n] = \frac{2f_1+f_1^2\tau^{\gamma}}{1+2f_1+f_1^2\tau^{\gamma}}\,\,.
\label{phit}
\end{equation}
These results are essentially identical to the free-energy minimization proposed by Aggeli et al \cite{peptides}. The only mathematical difference is that the LG model has an additional factor of $(1-\rho)(1-\rho_w)$ multiplying each cluster probability density $[n]$, to allow for arbitrarily high concentration. From the geometry of typical peptide tapes \cite{peptide2} one notes that a monomer concentration of $10^{-6}$ molar maps roughly to $\rho=10^{-6}$ in LG units. Since the experimental data for $\phi_t(\rho)$ in Fig.~6b of Aggeli et al is at concentrations less than 500 micromolar, this factor is completely irrelevant. Accordingly, we have identified an exactly solvable LG model with identical aggregation statistics to a known phenomenological analysis that has been successfully fitted to experimental data. The data imply a value of $\gamma$ around 0.2. There is however one physical difference with the free-energy analysis adopted in ref.\cite{peptides}; namely, instead of normalising by the volume of a monomer (as is automatic in the context of a 1D LG model) the authors choose a volume approximately 1000 times smaller. From (\ref{one},\ref{cdis}) this is equivalent to increasing the values of both $\tau$ and $\tau^{1+\gamma}$ by 1000, which is not sufficiently inconsistent at $\gamma=0.2$ to noticeably affect the fit to the data. However, it does result in a significant renormalization of the predicted binding energy $\ln\tau$ associated with the self-assembly. Aggeli et al describe their small volume as characterising the movement of monomers within a tape and thus might be regarded as a rough correction for the entropy cost associated with forming a linear aggregate in a 3D solution, in contrast to a 1D model.\\

By breaking the isodesmic symmetry with a non-zero value of $\gamma$ the self-assembly of dimers is suppressed to higher concentrations, after which the rest of the cluster distribution kicks in. This induces a sigmoidal shape to $\phi_t(\rho)$ just before the percentage of beta-sheet self-assembly reaches 50\%:
\begin{equation}
f_a=\sqrt{\tau^{\gamma}+1}-1\,\,\,\,;\phi_t=1/2\,\,.
\label{ctc}
\end{equation}
There is no discontinuous jump in the order parameter $\phi_t$ (no van der Waals `loop' in $\mu(\rho)$), since as already discussed there is no region of two-phase coexistence for these short-range 1D models. This behaviour is reminiscent of micelle formation; since the `micelles' are tapes, the concentration at which this rapid increase in linear aggregation takes place can be denoted a `ctc'. It is therefore of some interest that we have identified a class of exactly solvable 1D models that display the phenomenon of a ctc. These broken isodesmic LG models are defined explicitly by the hamiltonian
\begin{equation}
H = -(\epsilon+\epsilon_m)\sum_{i}t_it_{i+1}-\epsilon_m\sum_{i}t_i(1-t_{i-1})(1-t_{i+1})\,\,\,\,; t_i=0,1. 
\label{Hgam1}
\end{equation}
A non-zero value of $\gamma$ is therefore mathematically equivalent to the introduction of a three-body interaction.

\newpage

\newpage
\section*{FIGURE CAPTIONS}

\noindent FIG. 1. 
Insertion energies $\psi$ for evaluating the equation of state of a 1D non-isodesmic LG model from the PDT (\ref{pdt}). Single-hatched cells denote the constraint that a cell be empty, while double-hatched cells are constrained to be occupied. The insertion cell is constrained to be empty but is identified by leaving it unshaded. All remaining cells (not shown) have no constraints. Factors of 2X arise from 1D symmetry about the insertion cell.  
\\

\noindent FIG. 2. 
Insertion energies $\psi$ required when using the PDT (\ref{pdt}) to evaluate probabilities of occupancy directly to the right of constrained cells or walls, defining the cluster distribution of a 1D non-isodesmic LG model. (a) Empty wall: $\rho_w$; (b) Monomer wall: $\rho_1$; (c) Cluster wall: $\rho_a$.  Single-hatched cells denote the constraint that a cell be empty, while double-hatched cells are constrained to be occupied. The insertion cell is constrained to be empty but is identified by leaving it unshaded. All remaining cells (not shown) have no constraints.  

\newpage
\section*{CONTENTS PAGE}

$\,\,\,\,\,\,\,\,\,H = -(\epsilon+\epsilon_m)\sum_{i}t_it_{i+1}-\epsilon_m\sum_{i}t_i(1-t_{i-1})(1-t_{i+1})\,\,\,\,; t_i=0,1.$\\

\noindent
Broken isodesmic symmetry can be an important aspect of linear aggregation in solution. It is shown that experimental data on the self-assembly of peptide tapes can be mapped to exactly solvable one-dimensional models that incorporate this broken symmetry.

\begin{thebibliography}{99}
\bibitem{ising} E. Ising, Z. Physik {\bf 31}, 253 (1925).
\bibitem{PCagg} R. B. Martin, Chemical Reviews {\bf 96}, 3043 (1996). One argument is that the self-assembly of a dimer involves twice the loss of orientational entropy than when a monomer joins a cluster. 
\bibitem{peptides} A. Aggeli, M. Bell, N. Boden, J. N. Keen, T. C. B. McLeish, I. Nyrkova, S. E. Radford and A. Semenov, J. Mater. Chem. {\bf 7}, 1135 (1997). 
\bibitem{widom} B. Widom, J. Chem. Phys. {\bf 39}, 2808 (1963). In the current context (monomer insertion) this sum rule is best derived from enforcing detailed balance between the system of interest and an ideal-gas reservoir of monomers at the same temperature and chemical potential.
\bibitem{LL} L. D. Landau and E. M. Lifshitz, {\it Statistical Physics} 3rd Ed. Part I, (Pergamon, New York, 1980); final page.
\bibitem{peptide2} C. Whitehouse, J. Fang, A. Aggeli, M. Bell, R. Brydson, C. W. G. Fishwick, J. R. Henderson, C. M. Knobler, R. W. Owens, N. H. Thomson, D. A. Smith, and N. Boden, Angew. Chem. Int. Ed. {\bf 44}, 1965 (2005); see Fig.~2.
\end{thebibliography}
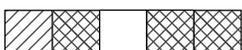
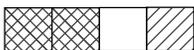
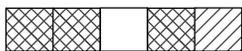
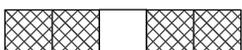
\end{document}